\documentstyle[aps]{revtex}
\tighten

\begin{document}
\draft
\title{Real-Time Dynamics from Imaginary-Time Quantum Monte Carlo
Simulations:\\ Tests on Oscillator Chains}
\author{J.~Bon\v ca\cite{byline} and J.E.~Gubernatis}
\address{Theoretical Division,
         Los Alamos National Laboratory,
         Los Alamos, NM 87545}
\date{\today}
\maketitle
\begin{abstract}
We used methods of Bayesian statistical inference and the principle of
maximum entropy to analytically continue imaginary-time Green's
function generated in quantum Monte Carlo simulations to obtain the
real-time Green's functions.  For test problems, we considered chains
of harmonic and anharmonic oscillators whose properties we simulated
by a hybrid path-integral quantum Monte Carlo method. From the
imaginary-time displacement-displacement Green's function, we first
obtained its spectral density.  For harmonic oscillators, we
demonstrated the peaks of this function were in the correct position
and their area satisfied a sum rule.  Additionally, as a function of
wavenumber, the peak positions followed the correct dispersion
relation. For a double-well oscillator, we demonstrated the peak
location correctly predicted the tunnel splitting.  Transforming
the spectral densities to real-time Green's functions, we conclude
that we can predict the real-time dynamics for length of times
corresponding to 5 to 10 times the natural period of the model. The
length of time was limited by an overbroadening of the peaks in the
spectral density caused by the simulation algorithm.
\end{abstract}

\section{Introduction}

One of the goals for doing computer simulations is the production of
information useful in the interpretation and design of experiments.
Notwithstanding important issues regarding Hamiltonian selection and
parameterization, the interface of simulations with experiment is
particularly challenging for quantum systems.  The current Monte Carlo
algorithms, whether they impose quantum particle statistics
constraints or not, are performed either in real-time $t$ or in
imaginary-time (Euclidean time) $\tau=-it$.  In real-time, the
propagator $\exp(itH)$ for a system, described by a Hamiltonian $H$,
oscillates wildly at long-times.  Analytically, these rapid
oscillations self-cancel, but a Monte Carlo process, as it is
typically used, has difficulty achieving this cancellation. As a
consequence, modifications of the basic algorithms have been proposed
to extend the simulations as long as possible in the real-time domain
\cite{berne86}.  With these new algorithms, simulations typically produce
dynamics extending to 2 to 3 times the natural periods of the systems.
In imaginary-time, the propagator $\exp(-\tau H)$ is diffusive and the
rapid oscillations are avoided.  Correlations functions $G(\tau)$,
however, are now a function of imaginary-time, and such functions do
not easily convey the actual dynamics of the system.  In principle,
real-time correlation (Green's) functions $\hat G(t)$ can be obtained
from the imaginary-time ones by the process of analytic continuation.
In practice, this process is difficult because it is ill-posed and
because the Monte Carlo data is incomplete and noisy.

Recently, procedures were proposed to perform this analytic
continuation \cite{gubernatis91}. These procedures draw heavily upon
methods of Bayesian statistical inference and the principle of maximum
entropy to infer from imaginary-time correlation functions their
associated spectral densities $A(\omega)$. Through linear-response
theory, the spectral densities represent the spectra associated with
numerous real-time measurements of current-current, spin-spin, etc.\
correlations functions. What apparently has not yet been tried is
performing the Hilbert transform of these spectral densities to obtain
the frequency-dependent retarded correlation function and then Fourier
transforming this quantity to obtain the real-time correlation
function.  In this paper, we will carry out these additional steps.
By doing this, we hope to gain a greater understanding of the physical
content present in the spectral density returned by the Bayesian
methods. We expected that the resultant real-time information would be
limited by the approximate and probabilistic nature of the analytic
continuation methods.  We found, however, that the distance in
real-time over which our results are valid is limited primarily by the
ability of the simulation algorithm to produce good data.  To
interface profitably with the numerical analytic continuation, the
simulation algorithm has to produce high quality data consistent with
the assumptions of procedures.  The algorithm we used had problems
doing this, and we will describe the measures taken to reduce this
difficulty. Even so, in most cases we were able to extend in real-time
up to factor of 10 natural periods of the physical systems.  Longer
extensions are possible and require longer Monte Carlo runs. For
present purposes, we had no physical motivation to do so.

In Section~II, we will discuss the various models studied. We
simulated a particle moving in single harmonic and double-well
anharmonic potential and a collection of particles moving in chains of
these potentials. For these models we know the exact solutions.  By
calculating their properties numerically, we can benchmark our
methods.  Certain properties of a single double-well potential, like
the tunnel splitting, are easily obtained numerically.  The phase
diagram for a chain of such oscillators is also known \cite{wang94}.  This
type of chain can exist in a symmetric or displacive state.  In
Section~III, we summarize the numerical analytic continuation
procedure we used and discuss our simulation technique. Modifying the
simulation technique to be more naturally ergodic and to produce data
with short statistical auto-correlation times was the most difficult
and restrictive part of our study.  We present our results and
conclusions in Sections~IV and V.

\section{Models}

We simulated five Hamiltonians. One was that for a single harmonic
oscillator
\begin{equation}
  H = \frac{p^2}{2m} + \frac{\gamma}{2} x^2
\label{eq:model1}
\end{equation}
which has the natural frequency $\omega_0=\sqrt{\gamma/m}$. Another was
that for a chain of $N$ such oscillators
\begin{equation}
  H = \sum_i \frac{p_i^2 }{2m}
           + \frac{\gamma}{2} (x_i-x_{i-1})^2
\label{eq:model2}
\end{equation}
Fourier transforming the displacements $x_i$ and momenta $p_i$, we can
of course rewrite this second Hamiltonian as
\begin{equation}
  H = \sum_k \frac{p_k^2}{2m} + \frac{1}{2}m\omega_k^2 x_k^2
\end{equation}
where $k=-\pi, -(N-1)\pi/N,\dots,\pi$ and $\omega_k^2 =
2\omega_0^2(1-\cos k)$.  In this form, the Hamiltonian is explicitly
expressed as a collection on $N$ independent simple oscillators. The
natural frequency of an oscillator is $\omega_k$. The third
Hamiltonian was a variant of the harmonic chain
\begin{equation}
  H = \sum_i \frac{p_i^2 }{2m}
           + \frac{\gamma}{2} (x_i-x_{i-1})^2 + \frac{1}{2} x_i^2
\label{eq:model3}
\end{equation}
which after Fourier transforming the displacements becomes
\begin{equation}
  H = \sum_k \frac{p_k^2}{2m} + \frac{1}{2}m\omega_k^2 x_k^2
\end{equation}
where $\omega_k^2 =
1+2\omega_0^2(1-\cos k)$.  In this form, the Hamiltonian is
again explicitly expressed as a collection on $N$ independent simple
oscillators but with a dispersion relation that has a non-zero
frequency at $k=0$. The two other Hamiltonians were a single,
symmetric, double-well potential
\begin{equation}
  H = \frac{p^2}{2m} + \frac{1}{4} (x^2-1)^2
\label{eq:model4}
\end{equation}
which has well bottoms at $x=\pm 1$ and a barrier height of unity at
$x=0$, and a chain of such potentials
\begin{equation}
  H = \epsilon \sum_i \frac{p_i^2}{2m}
                     +\frac{\gamma}{2} (x_i-x_{i-1})^2
                     +\frac{1}{4} (x_i^2-1)^2
\label{eq:model5}
\end{equation}
In the chain, $\epsilon$ sets the barrier height of every double-well.
In both chains we assumed periodic boundary conditions and disallowed
particle exchange.

For these Hamiltonians, our simulations obtained estimates of the
imaginary-time displacement-displacement Green's function
\begin{equation}
   G_{ij}(\tau) = G_{i-j}(\tau) = \langle T_\tau x_i(\tau)x_j(0)\rangle
\label{eq:gij}
\end{equation}
Here, the angular brackets denote thermal averaging.  It is more
convenient and illuminating to work with the spatial Fourier transform
$G_k(\tau)$
of $G_{ij}$ and it is known \cite{mahan81} that
\begin{equation}
  G_k(\tau) = \frac{1}{2\pi}\int_{-\infty}^\infty d\omega\,
\frac{e^{-\tau \omega} A_k(\omega)}{1-e^{-\beta\omega}}
\label{eq:the_equation}
\end{equation}
where $A_k(\omega)$ is the spectral density. This function has the
properties that
\begin{equation}
   A_k(\omega) = -A_k(-\omega)
\end{equation}
The odd symmetry of $A_k(\omega)$ allows us to re-express
(\ref{eq:the_equation}) as
\begin{equation}
  G_k(\tau) = \frac{1}{2\pi}\int_0^\infty d\omega\,
         \frac{\omega[e^{-\tau\omega}+e^{-(\beta-\tau)\omega}]}
              {1-e^{-\beta\omega}}
         \frac{A_k(\omega)}{\omega}
\label{eq:the_eq}
\end{equation}
and it is straightforward to show that $A_k(\omega)$ obeys the sum rule
\begin{equation}
  G_k(0) = \frac{1}{2\pi}\int_{-\infty}^\infty
           d\omega\, \frac{A_k(\omega)}{1-e^{-\beta\omega}}
         = \langle x_k^2\rangle
\label{eq:SumRule}
\end{equation}

The spectral density $A_k(\omega)$ is also related to the frequency
Fourier transform $\hat G^R(\omega)$ of the real-time, retarded Green's
function \cite{mahan81}
\begin{equation}
 \hat G_k^R(t)= -i\theta(t)\langle [x_{-k}(t),x_k(0)] \rangle
\end{equation}
via
\begin{equation}
   \hat G_k^R(\omega) = \frac{1}{2\pi}\int_{-\infty}^\infty
                d\omega'\, \frac{A_k(\omega')}{\omega-\omega'+i\eta}
\end{equation}
where $0 <\eta\ll 1$, from which it follows that
\begin{equation}
   \hat G_k^R(t) = -i\theta(t)\frac{1}{2\pi}\int_{-\infty}^\infty
                 d\omega\, A_k(\omega) e^{-i\omega t} e^{-\eta t}
\end{equation}

For an individual harmonic oscillator of frequency $\omega_k$,  the
eigenstates and energies are exactly known, and all
the quantities in the above paragraph are known analytically:
\begin{eqnarray}
  G_k(\tau) &=& \frac{1}{2m\omega_k}\frac{1}{\sinh(\beta\omega_k/2)}
            \cosh\left[\beta\omega_k
                 \left(\frac{1}{2}-\frac{\tau}{\beta}\right)\right]
                                                   \label{eq:GvsTau}\\
  A_k(\omega)&=& \pi
         [\delta(\omega-\omega_k)-\delta(\omega+\omega_k)]/m\omega_k
                                                   \label{eq:AvsOmega}\\
  \hat G_k^R(\omega) &=& \frac{1}{2m\omega_k}
               \left(\frac{1}{\omega-\omega_k+i\eta} -
                     \frac{1}{\omega+\omega_k+i\eta}\right)
                                                   \label{eq:GvsOmega}\\
  \hat G_k^R(t) &=& \theta(t)\frac{1}{m\omega}
                               \sin(\omega_kt) e^{-\eta t}
                                                   \label{eq:GvsT}\\
  \langle x_k^2\rangle &=& \frac{1}{2m\omega_k}
                           \coth\left(\beta\omega_k/2\right)
                                                    \label{eq:MeanSquare}
\end{eqnarray}

The eigenstates and energies of a single harmonic oscillator have
definite, well-known characteristics: Because the potential is
symmetric about $x=0$, the eigenfunctions have alternating parity.
The ground state has even parity and an energy $\frac{1}{2}\omega_k$.
The energies of the excited states are regularly spaced at intervals
of $\omega_k$.  The double-well potential is also symmetric about
$x=0$, and its eigenstates also alternate in parity with the ground
state again having even parity.  The precise details about the energy
spectrum, however, are only available numerically.  When these states
lie below the barrier, and particularly for deep wells, they group
into widely-separated, nearly degenerate pairs. The separation in
energies within and between pairs is called the {\it tunnel spitting}.
The spectral density is dominated by terms with matrix elements
involving states 0 and 1.  Contributions from matrix elements of $x$
involving (0,3), (3,4), (0,5), (3,5), (5,6), etc.\ have smaller
contribution to the spectral density because of smaller overlap
between the eigenstates.  Additionally, most can be ``frozen-out'' by
making the temperature at least comparable to $E_3-E_0$. This
temperature range is the one in which we generally worked.  The nature
of the energy levels and eigenstates is schematically represented in
Fig.~1.

The spatial Fourier transformation of the Hamiltonian of the
double-well chain does not produce as system of independent
oscillators. This is the essence of its non-linearity. The model,
however, has an interesting zero-temperature phase diagram as a
function of the parameters $\epsilon$ and $\gamma$ \cite{wang94}.
Roughly, $\epsilon$ is a measure of the barrier height relative to the
frequency of inter-site oscillation and the frequency of oscillation
associated with the well-bottom.  When the barrier height is large,
the particles collectively are displaced to the left or the right of
their classical equilibrium positions in a broken symmetry state
characterized by a non-zero value of the mean-squared displacement. As
the barrier height is lowered, a critical value is reached where
because of zero-point motion and tunneling, the particles collectively
make a transition to a state where the mean-squared displacement of
each is zero. Accordingly, a simulation of the chain, done in one of
these two thermodynamic phases, is expected to exhibit different
quantum characteristics in the spectral density.

\section{Methods}

\subsection{Hybrid Path-Integral Monte Carlo Method}

Our Monte Carlo simulations will be based on the Feynman path
integral formulation of quantum mechanics. In imaginary-time, this
formulation represents the partition function $Z$ as
\begin{equation}
  Z = \int {\cal D}x e^{-S[x(t)]}
\label{eq:partition}
\end{equation}
where
\begin{equation}
  S[x(\tau)] = \int_0^\beta d\tau\, H[x(\tau)]
\label{eq:action}
\end{equation}
is the action corresponding to the path $x(\tau)$ and $H[x(\tau)]$
represents the path dependence of the Hamiltonian. The Monte Carlo
method is used to perform the integration over the paths in
(\ref{eq:partition}). It is applied after the integral in
(\ref{eq:action}) is approximated by a sum over $L$ steps in
imaginary-time each of length $\Delta\tau$ and the momenta are
approximated by a forward-difference approximation between successive
displacements in imaginary-time:
\begin{equation}
  p_i(\tau) = m\frac{\partial x_i(\tau)}{\partial \tau}
  \approx m \frac{ x_i(\tau+\Delta\tau)- x_i(\tau)}{\Delta\tau}
\end{equation}
For a one-dimensional system of $N$ particles, the action becomes
\begin{equation}
 S = \Delta\tau\sum_{i,j=1}^{N,L}\frac{m}{2}
    (\frac{x_{i,j}-x_{i,j+1}}{\Delta\tau})^2
    +V(x_{i,j},x_{i+1,j})
\end{equation}
This form is similar in appearance to a classical two-body potential
energy defined on a $N\times L$ lattice where at a given $\tau$ the
particles interact through the potential energy function of the
original problem (scaled by $\Delta\tau$), and at a given position
they interact by a harmonic potential with a spring constant equal to
$m/\Delta\tau$.  For a single particle, the summation over the spatial
coordinate $i$ is dropped and the discretized action has the
interpretation of a chain where the particles at imaginary-time move
in a potential $\Delta\tau V(x_j)$ while interacting with particles at
neighboring times by a harmonic potential with spring constant
$1/\Delta\tau$.  For the models we are considering, the discretized
actions are:
\begin{enumerate}
\item Single Harmonic Oscillator:\\
\begin{equation}
S=\Delta\tau\sum_{j=1}^{L}
        {m\over 2}\left[{x_{j+1}-x_{j}\over\Delta\tau}\right]^2
       +{\gamma\over 2}x_{j}^2,
\end{equation}
\item Harmonic Chain:\\
\begin{equation}
S=\Delta\tau\sum_{i,j=1}^{N,L}
{m\over 2}\left[{x_{i,j+1}-x_{i,j}\over\Delta\tau}\right]^2
+{\gamma\over 2}\left(x_{i+1,j}-x_{i,j}\right)^2,
\end{equation}
\item Harmonic Chain plus on-site oscillator:\\
\begin{equation}
S=\Delta\tau\sum_{i,j=1}^{N,L}
{m\over 2}\left[{x_{i,j+1}-x_{i,j}\over\Delta\tau}\right]^2
+{\gamma\over 2}\left(x_{i+1,j}-x_{i,j}\right)^2+\frac{1}{2}x_{i,j}^2,
\end{equation}
\item Single Double-Well Potential:\\
\begin{equation}
S=\Delta\tau\sum_{j}^{L}
{m\over 2}\left[{x_{j+1}-x_{j}\over\Delta\tau}\right]^2
+{1\over 4}\left(x_{j}^2-1\right)^2.
\end{equation}
\item Double-Well Chain:\\
\begin{equation}
S=\epsilon\Delta\tau\sum_{i,j=1}^{L,N}
{m\over 2}\left[{x_{i+1,j}-x_{i,j}\over\Delta\tau}\right]^2
+{\gamma\over 2}\left(x_{i,j}-x_{i,j+1}\right)^2
+{1\over 4}\left(x_{i,j}^2-1\right)^2.
\end{equation}
\end{enumerate}

The simplest way to apply the Monte Carlo method is to move repeatedly
from point to point on the space-time lattice, at each point propose a
change in the coordinate, $x_{i,j} \rightarrow x_{i,j}'$, and accept
the change via the Metropolis algorithm with probability
$\min[1,\exp(-\Delta S)]$ where $\Delta S$ is the change in the value
of the action proposed by the proposed change \cite{creutz81}. This
method is often called {\it path-integral Monte Carlo} (PIMC).

An alternative to the Monte Carlo evaluation of the path-integral is a
molecular dynamics evaluation \cite{berne??}.  Here, a fictitious momentum
$\pi_{i,j}$ is associated with each point to define a pseudo
Hamiltonian
\begin{equation}
  H_p = \sum_{i,j=1}^{N,L} \frac{\pi_{i,j}^2}{2m_\pi} + S[x_{i,j}]
\end{equation}
and standard molecular dynamics techniques are used to sample phase
space.  The approach takes advantage of the classical nature of the
fields in the path-integral formulation and produces the correct
statistical mechanics because in classical statistical mechanics the
momentum degrees of freedom can be integrated out of the partition
function.  The method is often called {\it path-integral molecular
dynamics\/}.

At the core of the method we used is the hybrid Monte Carlo approach
suggested by Duane et al.\ \cite{duane87} which combines the molecular
dynamics approach with the Monte Carlo procedure to obtain the best
features of both methods.  The general expectation is faster
equilibration of the simulation and shorter auto-correlation times
between measured quantities. With this method, the following steps are
cycled: For a given set of $x_{i,j}$ the corresponding pseudo momenta
are assigned values randomly from a Maxwell-Boltzmann distribution for
the velocities.  The energy is computed. Next, both the momenta and
displacements are evolved by molecular dynamics for some pseudo time
$t_p$.  The energy is recomputed. Then the evolved displacements are
accepted with probability $\min[1,\exp(-\Delta E)]$, where $\Delta E$
is the difference in energy between the initial and final
configuration.  Normally, molecular dynamics is energy conserving so
the evolved displacements would always be accepted.  A guiding idea
behind the hybrid method is to use for the molecular dynamics with
emphasis on fast integration, as opposed to accurate integration, and
to adjust the size of these steps $\Delta t$ and $t_p$ so the Monte
Carlo decision accepts $90-95\%$ of the configuration.  The molecular
dynamics method globally updates all the displacements and is a
computationally efficient procedure.  The Monte Carlo procedure
maintains detailed balance to ensure proper equilibrium averages and
filters out the results of ``bad'' integrations.

We found, however, that this simple form of the hybrid method was
inadequate for present purposes.  The output of our simulation is to
be used as input to maximum entropy procedures to execute the analytic
continuation. As we will discuss below, the analytic continuation
problem is an ill-posed problem and hence is very sensitive to the
size of the errors associated with the input data.  For a fixed amount
of computer time, reducing the size of the error efficiently by a
Monte Carlo method requires shortening the auto-correlation times
between measurements. In our computed Green's function, in spite of small
estimates for the error bars, we would often see small (within the
error bars) unphysical sawtooth-like structure in regions about $\tau
= \beta/2$. Following a simple procedure suggested by Neal
\cite{neal95}, we could generally remove this structure and also be
more ergodic.  His suggestion was after each Monte Carlo decision to
reverse the direction of the molecular time integration, i.e., $\Delta t
\rightarrow -\Delta t$, with
probability $1/2$. A smaller improvement is achieved by not using a
fixed length for the time integration in the molecular dynamics
simulation but rather to choose the length randomly from the interval
$(t_p-\delta, t_p+\delta)$ where $t_p$ and $\delta$ are chosen so the
Monte Carlo acceptance rate is in the 90 to 95 per cent range.  We
will refer to the combined method as the {\it hybrid path-integral Monte
Carlo method} (HPIMC).  Adjusting the Monte Carlo acceptance rate is
not the entire story.  First, it seems best to insure $t_p$ is several
times larger than the natural period associated with the slowest
significant modes in the systems and then choose $\delta$ to fix the
acceptance ratio.

We remark that Fahy and Hamann \cite{fahy92} observed for the standard
hybrid method the existence of a critical time $t_c$ (dependent of
model parameters) demarking non-ergodic and chaotic behavior in the
results of the time integration.  For a harmonic system, $t_c$ is
infinite which suggests the inapplicability of the method to a
harmonic system.  We observed the behavior they found but whether
$t_p$ was larger or smaller than $t_c$ had only small consequences on
our measured results. As we illustrate below, we achieved
very accurate results for the harmonic models.

We also remark that the results of the simulations depend on the size
of $\Delta\tau$.  By performing simulations for several different
values of $\Delta\tau$, we could in principle extrapolate the results
to the $\Delta\tau=0$ limit.  We did not do this but instead observed
that simulations performed with different values of $\Delta\tau$ gave
very similar results.

\subsection{Maximum Entropy Method}

The maximum entropy method \cite{gubernatis91} is used to regularize
the solution of (\ref{eq:the_eq}). Dropping the subscript $k$ for
convenience, we rewrite this equations as
\begin{equation}
   G(\tau) = \int d\omega\, K(\tau,\omega) [A(\omega)/\omega]
\label{eq:the_integral_eq}
\end{equation}
where the kernel
\begin{equation}
  K(\tau) = \frac{1}{2\pi}
      \frac{\omega\bigl[e^{-\tau\beta}+e^{-(\beta-\tau)}\bigr]}
           {1-e^{-\beta\omega}}
\end{equation}
Because both the kernel and $A(\omega)/\omega$ are regular at
$\omega=0$, we solve for $A(\omega)/\omega$ and then trivially find
$A(\omega)$.  For discrete values of $\tau$ and $\omega$ we approximate
(\ref{eq:the_integral_eq}) as
\begin{equation}
   G_i = \sum_{i,j} K_{i,j} A_j
\label{eq:the_matrix_eq}
\end{equation}
where $G_i = G(\tau_i)$, $K_{i,j}= K(\tau_i,\omega_j)$, and $A_j =
A(\omega_j)\Delta\omega_j/\omega_j$.

The Monte Carlo method will return estimates $\bar G_i$ of $G_i$ and
estimates of the sample variance $\sigma_j^2$ on the $G_i$.  With this
information, a natural solution path to $A_j$ would be to find the
values of $A_j$ that minimize
\begin{eqnarray}
 \chi^2 &=&\sum_i(\bar G_i- G_i)^2/\sigma_i^2\\
        &=& \sum_i \Bigl(\frac{\bar G_i - \sum_j
                               K_{i,j}A_j}{\sigma_i}\Bigr)^2
\label{eq:chi_1}
\end{eqnarray}
This approach, however, almost always fails.  One reason is that it
ignores the strong correlations that normally exist between the
measured values of $G_i$, i.e., values of the Green's function at
different imaginary-times.  At the very least, we must modify
(\ref{eq:chi_1}) to be
\begin{equation}
   \chi^2 = \sum_{i,j}(\bar G_i- G_i) [C^{-1}]_{i,j}(\bar G_j - G_j)
\label{eq:least-square}
\end{equation}
where $C_{i,j}$ is the measured covariance among the values of $\bar
G_i$. The $i$-th diagonal element of $C$ is simply $\sigma_i^2$.  This
modification of the definition of $\chi^2$ while necessary is
insufficient.  The difficulty is the inverse problem, that is, solving
(\ref{eq:the_integral_eq}) for $A(\omega)$, is ill-posed.  This
condition is caused by the exponential character of the kernel at
large values of $\omega$.  At large $\omega$, large variations in
$A(\omega)$ make little change in $G(\tau)$.  The simulation, on other
other hand, gives noisy and incomplete information about $G(\tau)$,
and hence for a given set of $\bar G_i$, an infinite number of $A_j$
will satisfy the least-squared estimate (\ref{eq:least-square}).

The next level of solution seeks to regularize the minimization of
$\chi^2$ by constraining it, i.e., minimizing
\begin{equation}
    \chi^2 -\sum_i \alpha_i f_i(A)
\end{equation}
where the $\alpha_i$ are Lagrange multipliers and the $f_i(A)$ are
functions of $A_j$ representing possible constraints on the solution.
Typical constraints include smoothness, non-negativity, sum rules,
moments, etc.  The difficulty with this approach is choosing the
Lagrange multipliers.  {\it Ad hoc} choices are commonplace.  Often small
changes in the values of these parameters produce massive changes in the
results.

The maximum entropy approach follows from the observation that the
spectral density is interpretable as a probability density function.
The principle of maximum entropy states the probabilities should be
assigned in such a way as to maximize
\begin{equation}
  S= \sum_j A_j-m_j-A_j\ln(A_j/m_j)
\end{equation}
Here, the $m_j$, called the {\it default model}, set the location of
the maximum of $S$ and the value of $S$ at this point to be zero.  The
default model is solution for $A_j$ in the absence of other
constraints on $A_j$. The method of maximum entropy maximizes
\begin{equation}
  Q(A) = \alpha S - \frac{1}{2} \chi^2
\end{equation}
To fix $\alpha$, an ad hoc procedure called {\it historic} maximum
entropy is often used \cite{skilling89,gull89}.  An more modern
alternative is the Bayesian-based {\it classic} maximum entropy which
uniquely determines $\alpha$ provided certain conditions are meet
\cite{skilling89,gull89}.  Under these conditions solution for $A_j$
is the most probable one. Unfortunately, these conditions seem often
violated in the analytic condition problem. Accordingly, to estimate
the $A_j$, we adopted a procedure suggested by Bryan.

In {\it Bryan's method} \cite{bryan90}, for a given value of $\alpha$,
we find the $A(\alpha)$ that maximizes $Q(A)$.  For the solution to
(\ref{eq:the_matrix_eq}), we take
\begin{equation}
   \bar A = \int d\alpha\, A(\alpha)\Pr[\alpha|\bar G]
\label{eq:the_solution}
\end{equation}
where $\Pr[\alpha|\bar G]$ is the probability of $\alpha$ given the
data $\bar G$. Bayesian analysis shows that
\begin{equation}
   \Pr[\alpha|\bar G] = \Pr[\alpha]\int {\cal D}A\, \frac{e^Q}{Z_LZ_S(\alpha)}
\end{equation}
where $Z_L$ is the normalization factor for $e^{-\frac{1}{2}\chi^2}$,
$Z_S(\alpha)$ is the normalization factor for $e^{\alpha S}$, and
$\Pr[\alpha]$ is Jeffreys' prior.  Details on the computation of this
joint probability are discussed elsewhere \cite{gubernatis91}.  With
this function, the integral (\ref{eq:the_solution}) is performed
numerically.

The most difficult part of the problem is not evaluating the maximum
entropy equations but satisfying the statistical assumptions on which
they are based.  The principal assumption is
\begin{equation}
   \Pr[\bar G|A]= \frac{e^{-\frac{1}{2}\chi^2}}{Z_L}
\label{eq:likelihood}
\end{equation}
The meaning of this assumption is the measured values of $G_i$ are
statistically independent and distributed according to a
multi-variable Gaussian distribution function defined by the
covariance matrix $C$.  Proper estimation of $C$ is paramount. Under
normal circumstance the data produced by the simulations do not
satisfy these assumptions. The procedures we use to have the data
approximate these assumptions are discussed elsewhere
\cite{gubernatis91}.  When we have proper data, our solution
(\ref{eq:the_solution}) usually shows good insensitivity to the choice
of the default model.  Additionally, the historic and classic maximum
entropy solutions usually agree well with it.

\section{Results}

To determine spectral properties of models listed in Section II, we
performed HPIMC simulations with up to 800 bins of data ($N_{bin}$) ,
each with up
to 4000 measurements ($N_{sweep}$).  Simulations with large bin-sizes were
necessary
to avoid nonergodic behavior of the HPIMC method when used for chains
with double well potentials close to the zero-temperature phase
transition point.  Furthermore, we set the value for the imaginary
time step to $\Delta\tau=0.25$. This choice on the one hand was small
enough to avoid errors associated by the discretization of otherwise
continuous imaginary time scale $\tau$, and on the other hand was
large enough to avoid unphysical correlations between successive
imaginary-time measurements of the Green's function $G(\tau)$. Since
our calculations were performed at the inverse temperatures $\beta$
between 1 and 10, the corresponding number of imaginary time steps
$L=\beta\Delta\tau$ was between 40 and 60.

For a successful application of the HPIMC it is crucial to choose the
proper value of the pseudo-time $t_p$ and the size of its step $\Delta
t$ in the molecular dynamic part of the simulation. Following Fahy and
Hamman, we determined the value of the critical value $t_c$ for each
case under consideration and then took $t_p>t_c$ to avoid running the
simulation in a non-ergodic regime. Typical values for $t_p$ were
between 5 and 15 for double-well cases listed below. We stress that
there was not much difference in the quality of the HPIMC data if we
chose $t_p>t_c$ or chose $t_p = t_c/2$. In addition, we obtained good
data for the harmonic wells by choosing $t_p\sim 1/\omega_0$ even
though for this particular case $t_c=\infty$. If we define $\omega_0$
as the smallest nonzero frequency of the system, optimal values of the
step size $\Delta t$ are between $0.05/\omega_0$ and $0.1/\omega_0$.
Larger values of $\Delta t$ lead to larger errors in the pseudo-time
propagation which then lead to small acceptance ratios. Smaller
values of $\Delta t$ lead to longer compution times.
Unless specified otherwise, we always chose $m=m_\pi=\gamma=1$. We
emphasize that the HPIMC method is insensitive of the choice of the
mass $m_\pi$ associated with the fictitious momentum $\pi_{i,j}$.

In Fig.~2, we show the displacement-displacement Green's function
$G(\tau)$, for a single harmonic oscillator, obtained by evaluating
(\ref{eq:GvsTau}) for various values of $\omega_0\beta$, as a function
of the imaginary-time variable $\tau$.  These curves look similar to
the Green's functions that we obtained numerically for this and the
other models: For some parameters and temperatures, the $G(\tau)$
varies little as a function of $\tau$; for others, it varies
rapidly at the ends of the interval $[0,\beta)$ and is flat
in the middle with values nearly equal to zero; and for still
other parameters, it has a featureless, parabolic looking shape.

The common features of these curves have several significant
implications for the analytic continuation problem. First, we remark
that from the quantum Monte Carlo simulations we obtain estimates of
$G(\tau)$ only at a relatively small number of discrete values
$\tau_i$.  The smoothness of the curves implies that the $G(\tau_i)$
at neighboring values of $\tau_i$ are correlated.  The computation of
the covariance matrix in (\ref{eq:least-square}) is thus an important
part of the of the analysis of the data.  While the correlations among
the different $\tau$ values of $G(\tau)$ make the interpretation of
the assignment of an ``error-bar'' to a given $\tau$ value delicate,
such an assignment illustrates several difficulties inherent in the
data that help to make the analytic continuation of the data often
very difficult.  In the case where $G(\tau)$ is nearly flat, the
errors bars mean that a number of values of $G(\tau)$ are ``within the
error-bars'' of each other.  This situation, along with the
correlations implied by sizable off-diagonal elements of the
covariance matrix, means that only a subset, and often a small subset,
of the measured values of $G(\tau)$ represent independent data useful
for the analytic continuation procedure.  The analytic continuation
near the classical limit can be very difficult.

The situation with the rapid end-point variation and the flat nearly
zero values is another difficult case. Again the flat region generates
a lose of useful values of $G(\tau)$ and the smallness of $G(\tau)$ in
this region can engender situations where the error bars would imply
that during the simulation estimates of $G(\tau)$, which must be
non-negative, were derived from ensemble values that included negative
ones. The Monte Carlo algorithms in fact do not produce negative
values but do produce highly skewed fluctuations about the mean.  The
Gaussian assumption for the likelihood function in
(\ref{eq:likelihood}) thus can only be approached in the limits of a
large number of independent measurements when the central limit
controls the data distribution. The ratio of the
mean value to the estimated variance (signal to noise ratio) also
indicates that the most effective data comes from those in the rapidly
decreasing region.  At low temperatures, the analytic
continuation problem can become very difficult.

The details of the simulation algorithm can also impact the quality of
the results and data.  In Fig.~3, we show $G(\tau)/G(0)$ as a function
of $\tau$ obtained by two closely related simulation techniques for a
single harmonic oscillator. We remark that the scale of abscissa is
$1/100$ of that of Fig.~2 and the ordinate shows $\tau$ only in a
narrow region at the symmetry point $\beta/2$. The dashed curve is the
analytic result obtained from (\ref{eq:GvsTau}).  The data is
represented by square markers were obtained by the HPIMC method
without the time-reversed step; the data represented by the circles
were obtained with the HPIMC method with the time-reversed step.  In
each case, the same number of Monte Carlo steps were made.  One sees
that the fluctuations with the HPIMC method without the time-reversed
step are larger and that the error bars associated with the results
suggest a dip into non-negative values of $G(\tau)$.  More
significantly, the results deviate from the exact curve by more than
one standard deviation in the immediate vicinity of $\tau=\beta/2$.

The analytic continuation result for $A(\omega)$ from the data
partially shown in Fig.~3 is shown in Fig.~4.  From
(\ref{eq:AvsOmega}), the spectral density should be
$0.5\delta(\omega-1)$.  The solid curve is obtained from the HPIMC
algorithm and shows a broadened delta-function at the right location
with nearly the correct weight. The fraction of a per-cent difference
from the correct weight is most likely a consequence of the small
error caused by discretizing the imaginary-time derivatives.  On the
other hand, the dashed curve, which is obtained from the data obtained
from the HIPMC algorithm without the time-reversed step, is broader,
located incorrectly, and has a larger discrepancy in its weight.  The
increased breadth is a consequence of the larger variance in the
measured data.  The incorrect location and poorer weight is a
consequence of the small deviation from the exact value near
$\tau=\beta/2$, which in turn is a consequence of the Monte Carlo
algorithm performing badly.

In Fig.~5, with the data obtained from the HPIMC algorithm, we show
the full analytic continuation form $G(\tau)$ to $G^R(t)$.  In
Fig.~5a, the data (open circle) is compared with the exact results
for $G(\tau)$ (solid line) obtained from (\ref{eq:GvsTau}).  In
Fig.~5b, the dashed curve is the $A(\omega)$ obtained by the analytic
continuation procedure, while the solid line is a Lorentzian at the
same location.  The real part of $G(\omega)$ is shown in Fig.~5c,
where the dashed line is the quantum Monte Carlo results and the solid
line is an analytic result obtained from the Lorentzian from Fig.~5c.
The width $\eta$ of the Lorentzian shown in Fig.~5b was adjusted so
the $\omega=0$ values of the two curves agreed.  The single
adjustment produced remarkably good agreement at high frequencies.
The principal differences between the two curves are at $\omega=\pm 1$
where divergences should exist as indicated by (\ref{eq:GvsOmega}).
Finally, $G^R(t)$ is shown in Fig.~5d.  The solid line was obtained
analytically and used the Lorentzian of Fig.~5b, while the dashed line
is the quantum Monte Carlo result.  The agreement between the results
is satisfying and comparable in quality to that possible by real-time
quantum Monte Carlo methods.  The width of the Lorentzian was
$7/\omega_0$, i.e., about 7 times the natural period of the harmonic
oscillator.  This width is still controlled by the size of the
variance in the measured values of $G(\tau)$.

For the case of the harmonic chain, we computed the Green's function
$G_k(\tau)$ for each independent wave number and from it found the
corresponding spectral density $A_k(\omega)$.  As with the single
oscillator, this function should be a delta-function located at
$\omega_k$ and have a weight equal to $G_k(0)=\langle x_k^2 \rangle$.
As in the single oscillator case, instead of finding a delta-function,
we found a Lorentzian-like peak at the correct location with the
correct weight.  The peaks for different values of $k$, however, had
different widths.  In general, the peak widths increased with
increasing $k$, correlating with the increased variance associated
with the $G_k(\tau)$. The spectral densities for the three lowest
values of $k$ are shown in Fig.~6.  The peak positions as a function
of $\omega$ give the phonon dispersion relation.  Our determination of
this relation is compared to the exact result in Fig.~7.  The
agreement is excellent.  There is a difficulty that must be mentioned.
At $k=0$ and $\omega_k=0$, $G_k(\tau)$ is flat and the weight of
the peak approaches infinity.  For $k=0$, not surprisingly, we were
unable to do the analytic continuation.  This situation is why we
considered the model defined by (\ref{eq:model3}).  Here, $\omega_k$
at $k=0$ is not zero and the determination of $A_k(\omega)$, and
subsequently $\omega_k$, is possible for all values of $k$.  The
dispersion relation found from the analytic continuation agrees very
well with exact result.

For a single, double-well potential, the spectral density can give
direct information about tunneling processes. This situation is
illustrated in Fig.~8 where the spectral densities for the
Hamiltonian, described by (\ref{eq:model4}), are shown for several
different values of the parameter $\gamma$.  It is straightforward to
discretize Schroedinger's time-independent differential equation for
this potential and find the eigenvalues of the resulting eigenvalue
equation. We adjusted the model parameters so only a very few (usually
one) of the lowest eigenstate lied below the barrier height,
similar to the situation depicted in Fig.~1a.  (For deeper wells, our
simulation methods would get stuck in one well or the other for large
numbers of Monte Carlo steps, and hence the algorithm effectively lost
ergodicity.)  In the cases reported, the temperatures of the
simulations were also less than the separation between the two lowest
lying pairs of eigenstates.  Thus, our spectral densities only
exhibited the transition between these two states, and the position of
the peak gives a direct measure of the lowest frequency tunnel
splitting.  This position agreed very well with the exact value
calculated from Schroedinger's equation.  Additionally, the weight of
the peak also agreed well with the sum rule (\ref{eq:SumRule}).

For a chain of double-well potentials (\ref{eq:model5}), still a
different situation presents itself.  At zero temperature, depending
on the values of $\epsilon$ and $\gamma$, the system exists in either
a broken symmetry phase, in which the mean-squared displacement is
non-zero, or in a symmetric phase, in which the mean-squared
displacement is zero.  The model has a quantum phase transition. We
found that simulation, if performed in the broken symmetry phase,
stuck in one well or the other for large numbers of Monte Carlo
steps, making it difficult to collect the amount of statistically
independent information to do the analytic continuation.  In Fig.~9,
we report the $A_k(\omega)$ for two simulations done in the symmetric
phase. The one in Fig.~9b is close to the zero-temperature phase
boundary. As one moves closer to the $T=0$ phase boundary by increasing
$\epsilon$, the $k=0$ peak moves towards $\omega=0$. This movement is a
consequence of the decreasing probability for tunneling between the two
minima of the double-well potential as the barrier height is
increased.

\section{Concluding Remarks}

We used methods of Bayesian statistical inference and the principle of
maximum entropy to analytically continue imaginary-time Green's
function generated in quantum Monte Carlo simulations to obtain the
real-time displacement-displacement Green's functions.  For test
problems, we considered chains of harmonic and anharmonic oscillators
whose properties we simulated by a hybrid path-integral quantum Monte
Carlo method (HPIMC). From the imaginary-time,
displacement-displacement Green's function, we first obtained its
spectral density.  For harmonic oscillators, we demonstrated that the
peaks of this function were in the correct position and their area
satisfied a sum rule.  Additionally, as a function of wavenumber, the
peak positions followed the correct dispersion relation. For a
double-well oscillator, we demonstrated the peak location correctly
predicted the tunnel splitting.  Transforming the spectral densities
to real-time Green's functions, we conclude that we can predict the
real-time dynamics for length of times corresponding to 5 to 10 times
the natural period of the model. The length of time was limited by the
simulation algorithm and not directly by the analytic continuation
procedure.

The simulation algorithm influences the results in at least two ways.
One way is the ease with which statistically independent,
Gaussian-distributed data is obtained.  In our experience, of the
various quantum Monte Carlo methods we have used, path-integral Monte
Carlo methods (PIMC) tend to produce data with long-ranged
correlations, thus making statistical independence sometimes difficult
to achieve. Achieving statistical independence is important for proper
error estimation because of the sensitivity of ill-posed problems to
errors in the data.  With out statistical indepence these error are
usually underestimated. That the data is Gaussian-distributed is
necessary to satisfy the assumptions of the analytic continuation
procedure.  Currently, large amounts of data are used to force the
simulation data to have proper statistical properties.  The method of
binned averages \cite{gubernatis91}, with many measurements in a bin,
is used to achieve statistical independence.  The central limit
theorem is used to obtain a Gaussian distribution of the binned averages.
Because of the low computational intensity necessary for the test
cases considers, producing the necessary large amounts of data was not
problematic.

A second way the algorithm can influence the results is through broken
ergodicity and unphysical results.  Here, we are referring to the
ragged structure we saw in $G(\tau)$ with the hybrid method and to the
small systematic difference between the exact and computed results.
We have not previously seen similar problems.  Small modifications in
the simulation algorithm removed the problems but it was at first
difficult to determine the source of the difficulty.  Because the
purpose of the research was not algorithmic development, we did not do
any comparisons of HPIMC and other possible methods to determine
relative efficiency and other merits.  Again, the computation times
for these simulations is small (a few hours on a modest workstation)
and so we were unmotivated to make such comparisons.  Other recently
suggested approaches, e.g., \cite{frantz92,cao}, should be considered
as part of further studies.

In general, we believe we have demonstrated that analytically
continuing imaginary-time correlation functions, obtained from a
quantum Monte Carlo simulation, to obtain real-time correlation
functions is a feasible alternative to obtaining such real-time
functions directly from a quantum Monte Carlo simulation done in
real-time.  One advantage in choosing this approach appears to be the
longer length of time over which the real-time information is faithful
to the correct result.  Of course, this conclusion is based on one
study of simple models; however, this study does strongly indicate the
direction for further and more extensive work.

\acknowledgments

This work was supported by the U.S. Department of Energy.  We thank J.
Doll and G. Berman for helpful conversations.

\newpage


%
%

\newpage

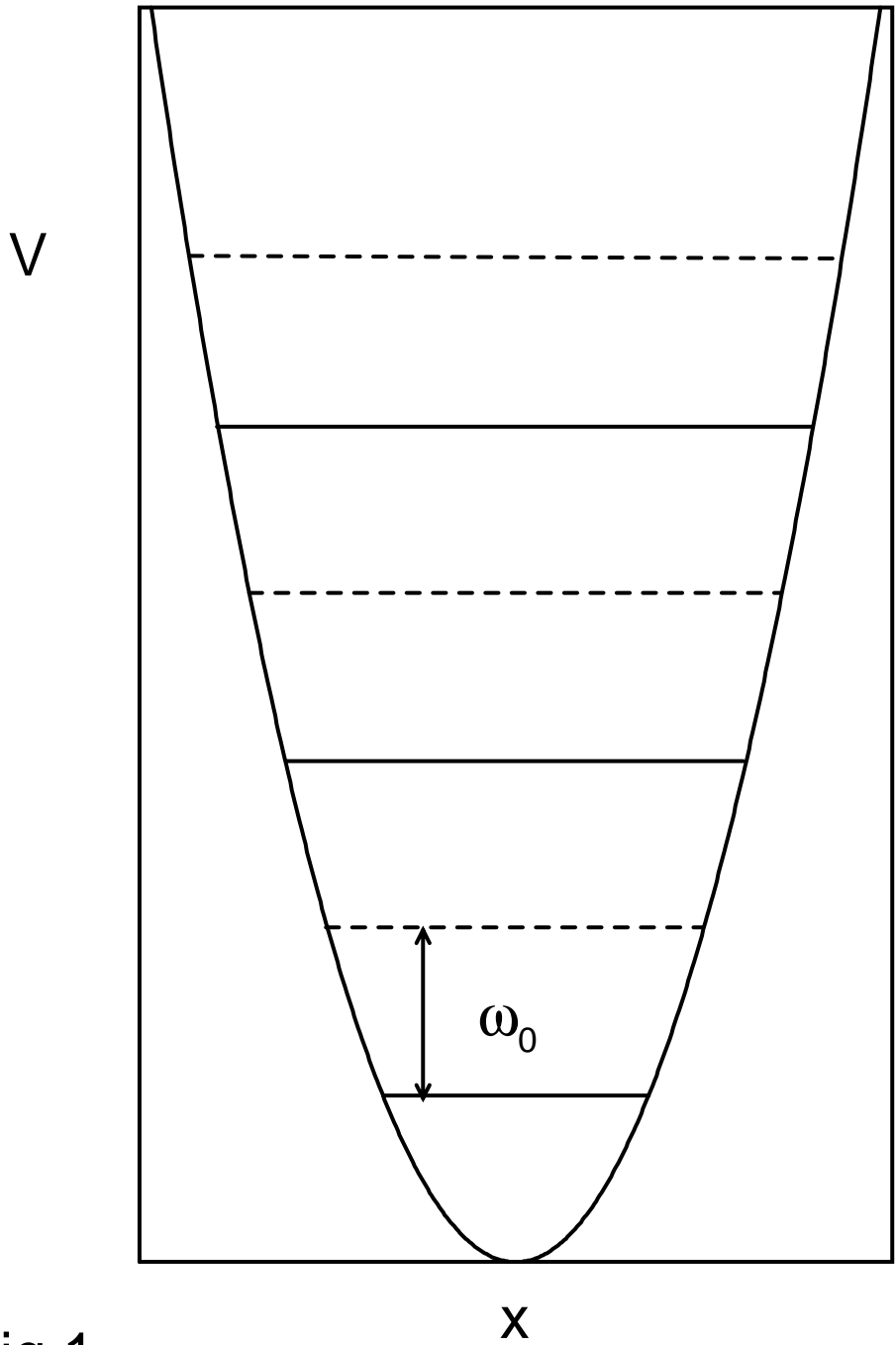
\begin{figure}
\caption[]{ Schematic representation of the energy levels in a
harmonic and double-well potential.}
\label{fig1}
\end{figure}

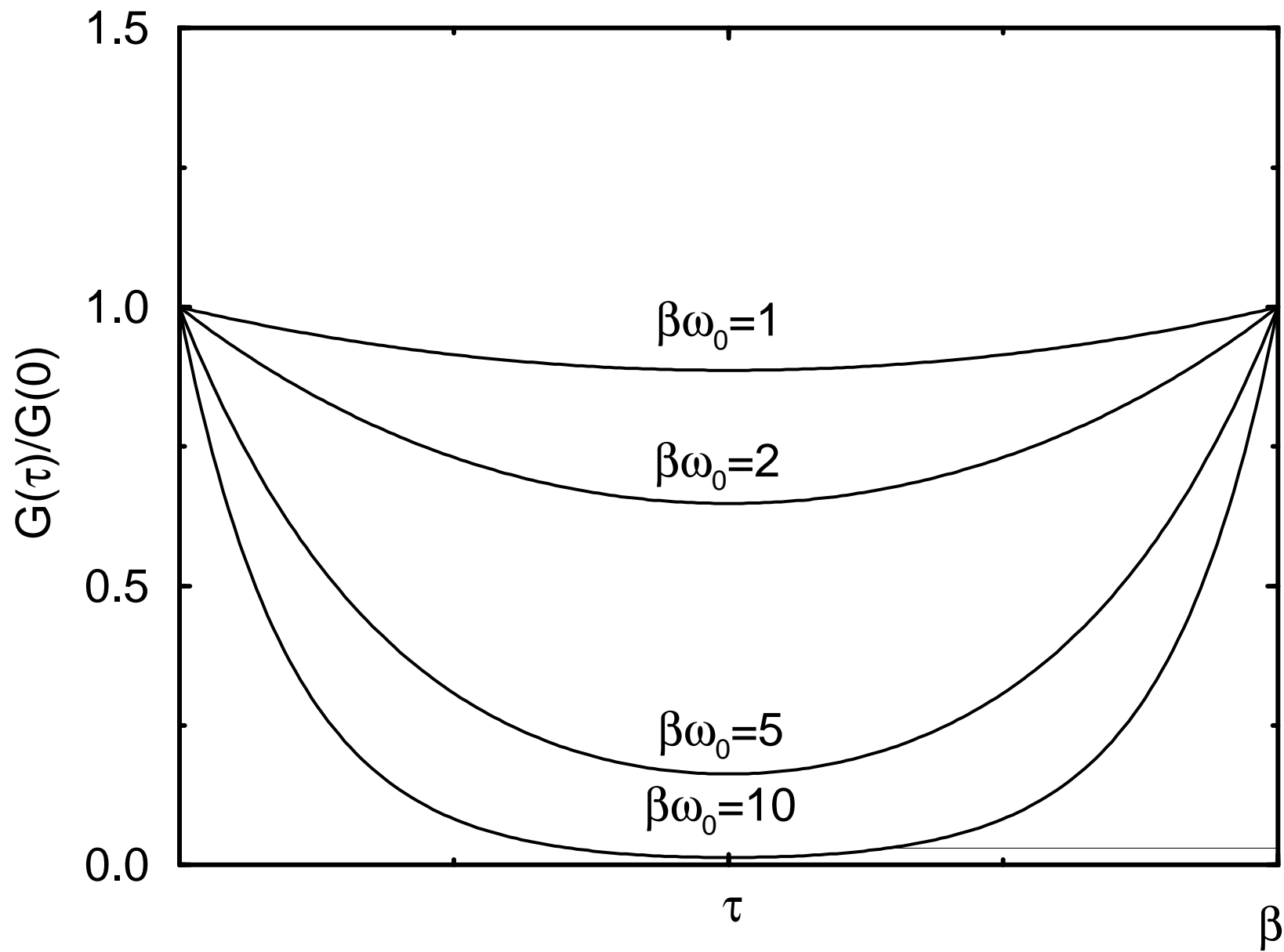
\begin{figure}
\caption[]{Green's function $G(\tau)$ for a particle in a
harmonic-well, obtained from the analytical form (21) potential, as a
function of imaginary-time $\tau$ and at different values of
$\beta\omega_0$.}
\label{fig2}
\end{figure}

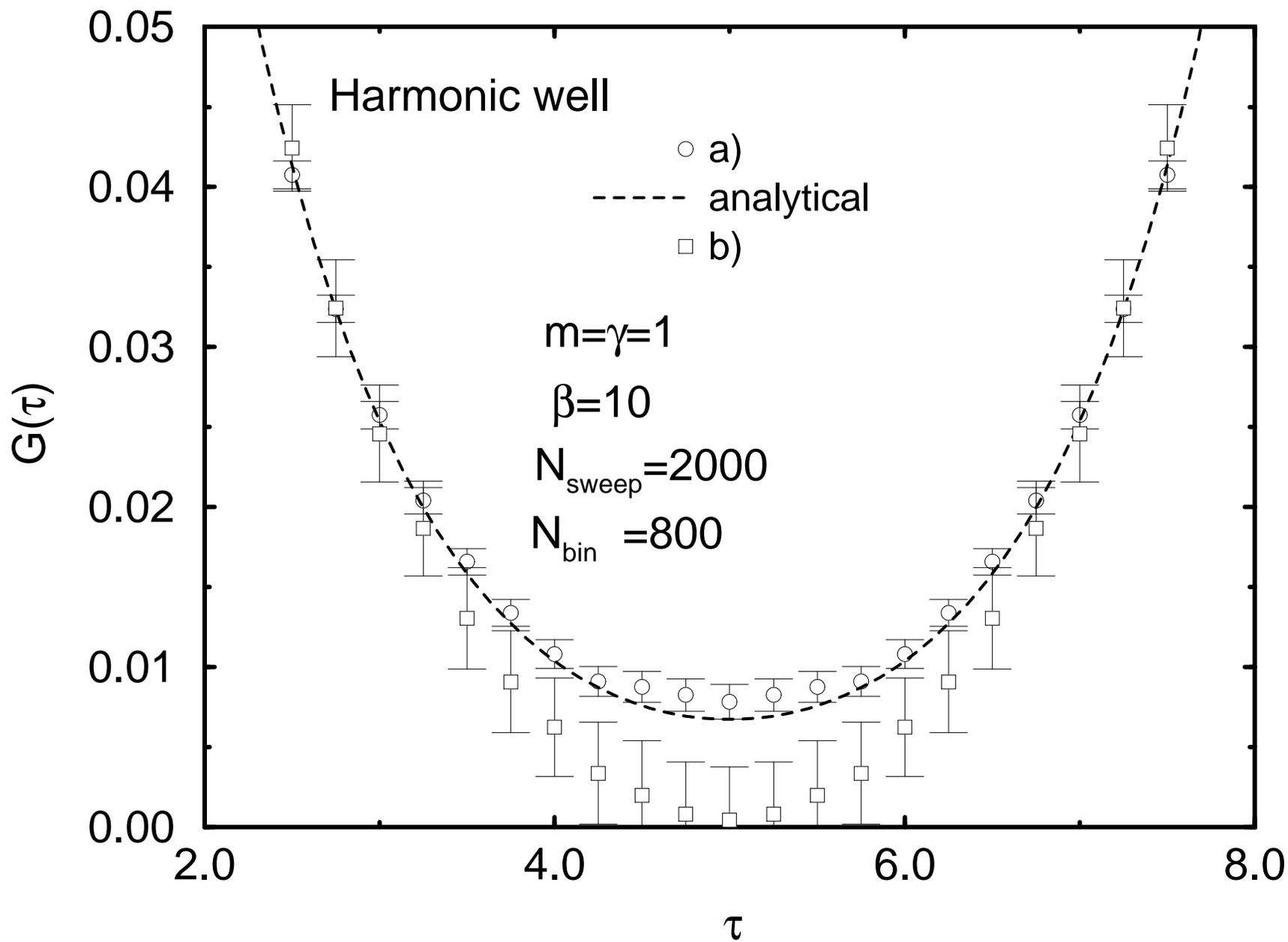
\begin{figure}
\caption[]{Comparison of analytical results for the Green's functions for
a particle in an harmonic well (dashed curve) with numerical results
(open circles and squares) obtained using HPIMC method (a) with and (b)
without the time-reversal step in the molecular dynamic part of the
algorithm.}
\label{fig3}
\end{figure}

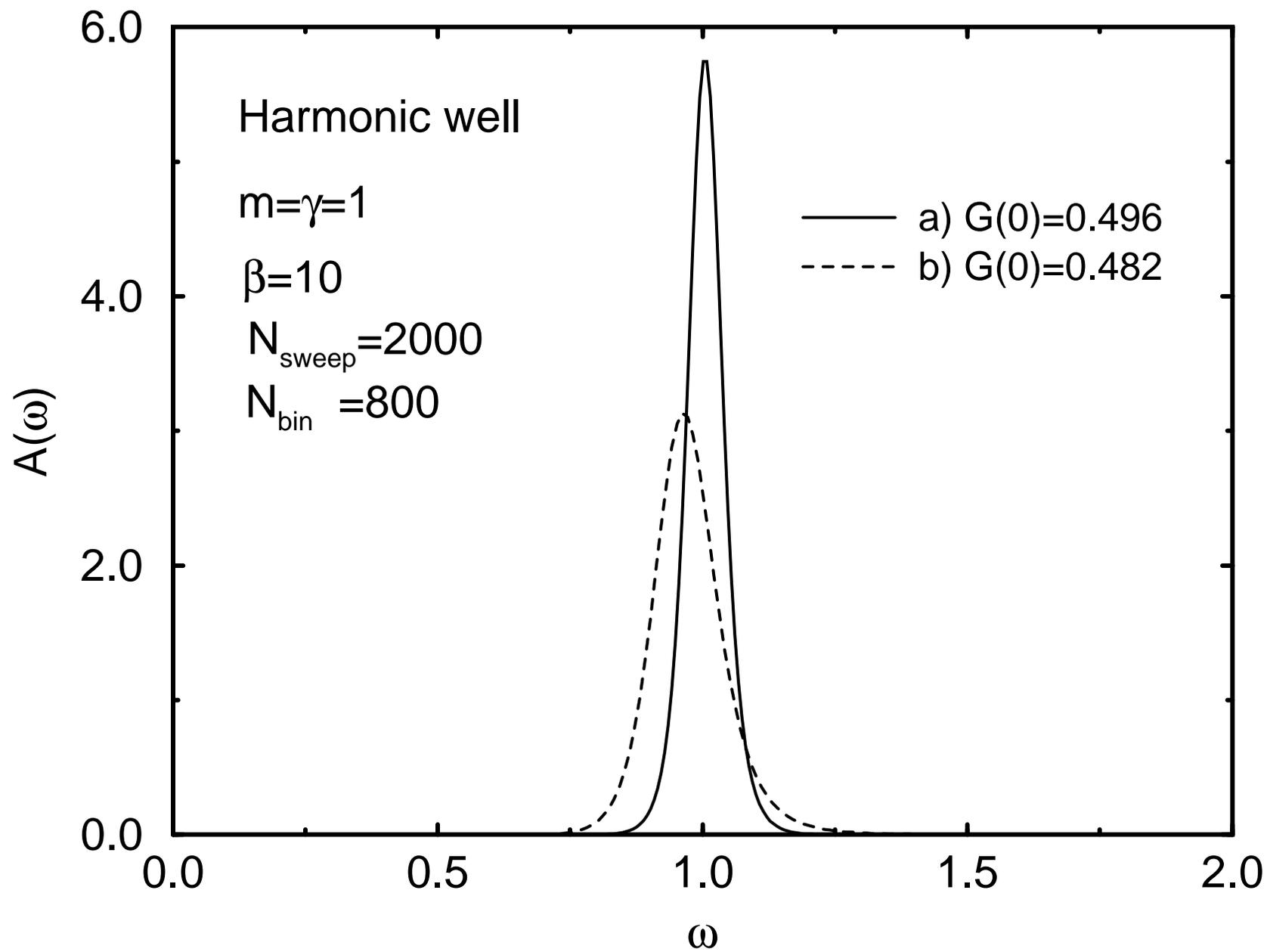
\begin{figure}
\caption[]{Comparison of the spectral functions obtained from the data
shown in Fig.~3. As in Fig.~3, in we present spectral functions
extracted from HPIMC data (a) with and (b) without the time-reversal
step. We also present $G(0)$, the computed area under the curves.}
\label{fig4}
\end{figure}

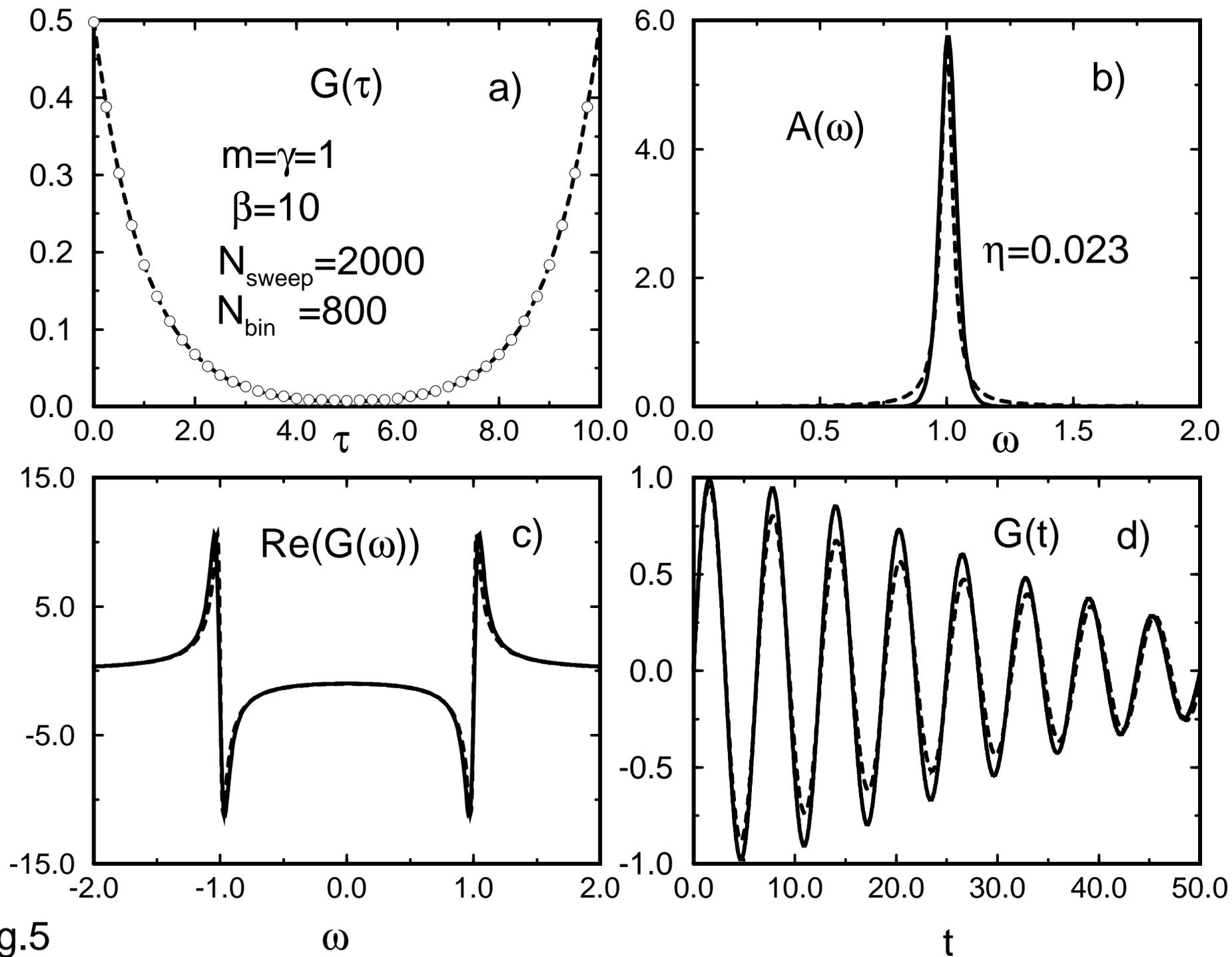
\begin{figure}
\caption{Comparison of the Green's function (a) and its spectral
fruntions for a particle in a harmonic well obtained by randomly
changing the direction of the molecular dynamics time integration
(open circles in (a) and full line in (b)) and by not randomly
changing he integration time direction (open circles in (a) and full
line in (b)). Presented is only a small portion of $G(\tau)$ around
$\tau=\beta/2$ where deviations from the analytic solution (dashed line
in (a)) are enhanced.  Also presented are the sum rules. Analytically,
$G(\tau=0)=\langle x^2 \rangle =0.500$.}
\label{fig5}
\end{figure}

\begin{figure}
\caption[]{Spectral functions for the harmonic chain at different
values of $k$. Arrows above spectral functions denote exact positions of
peaks. We also compare the numerically calculated sum-rules $f$ with
analytical values.}
\label{fig6}
\end{figure}

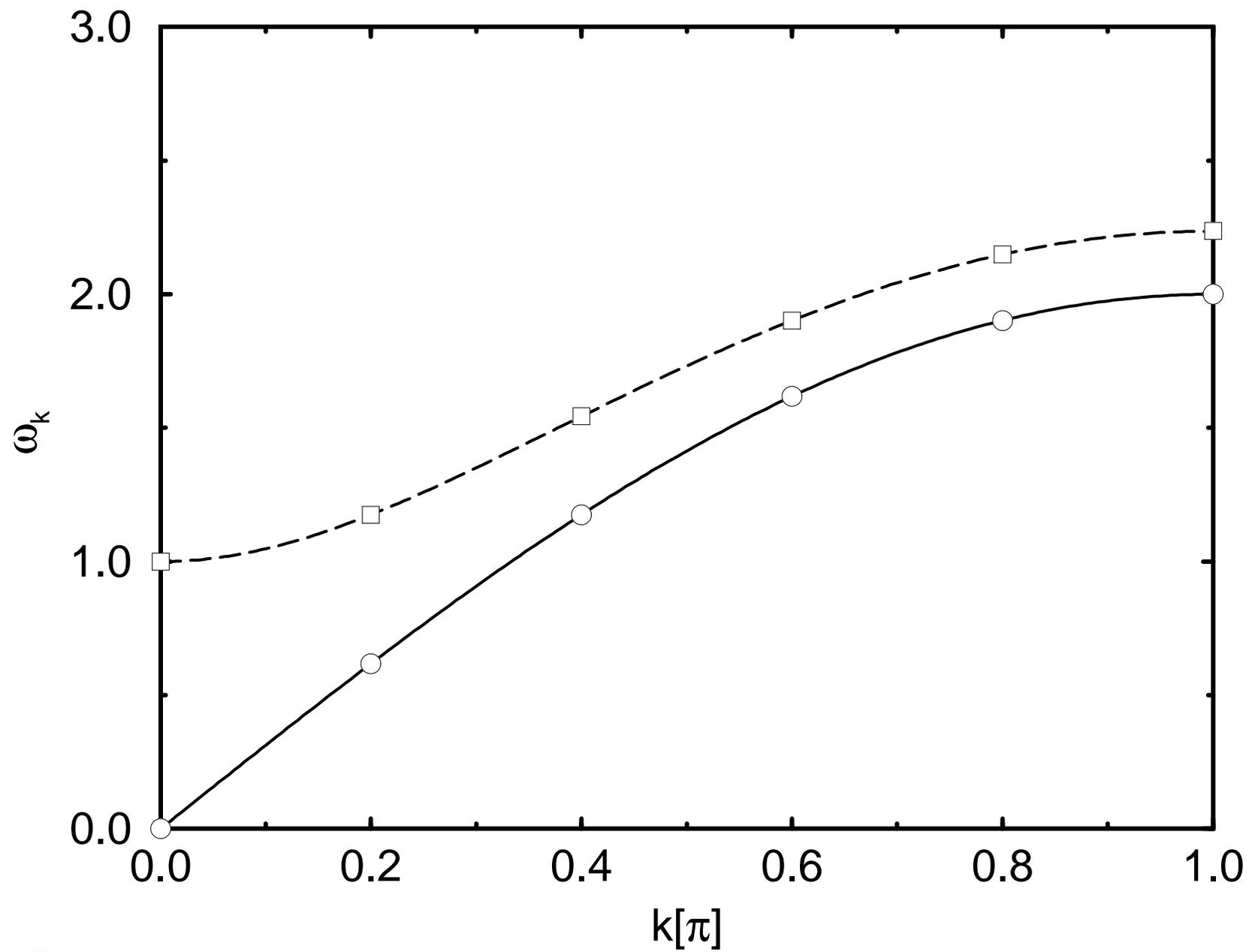
\begin{figure}
\caption[]{Dispersion relation $\omega_k$ for harmonic chain with
$m=1$ and $\gamma = 1$. Open circles correspond to discrete values of
$\omega_k$ for a 10-site system.}
\label{fig7}
\end{figure}

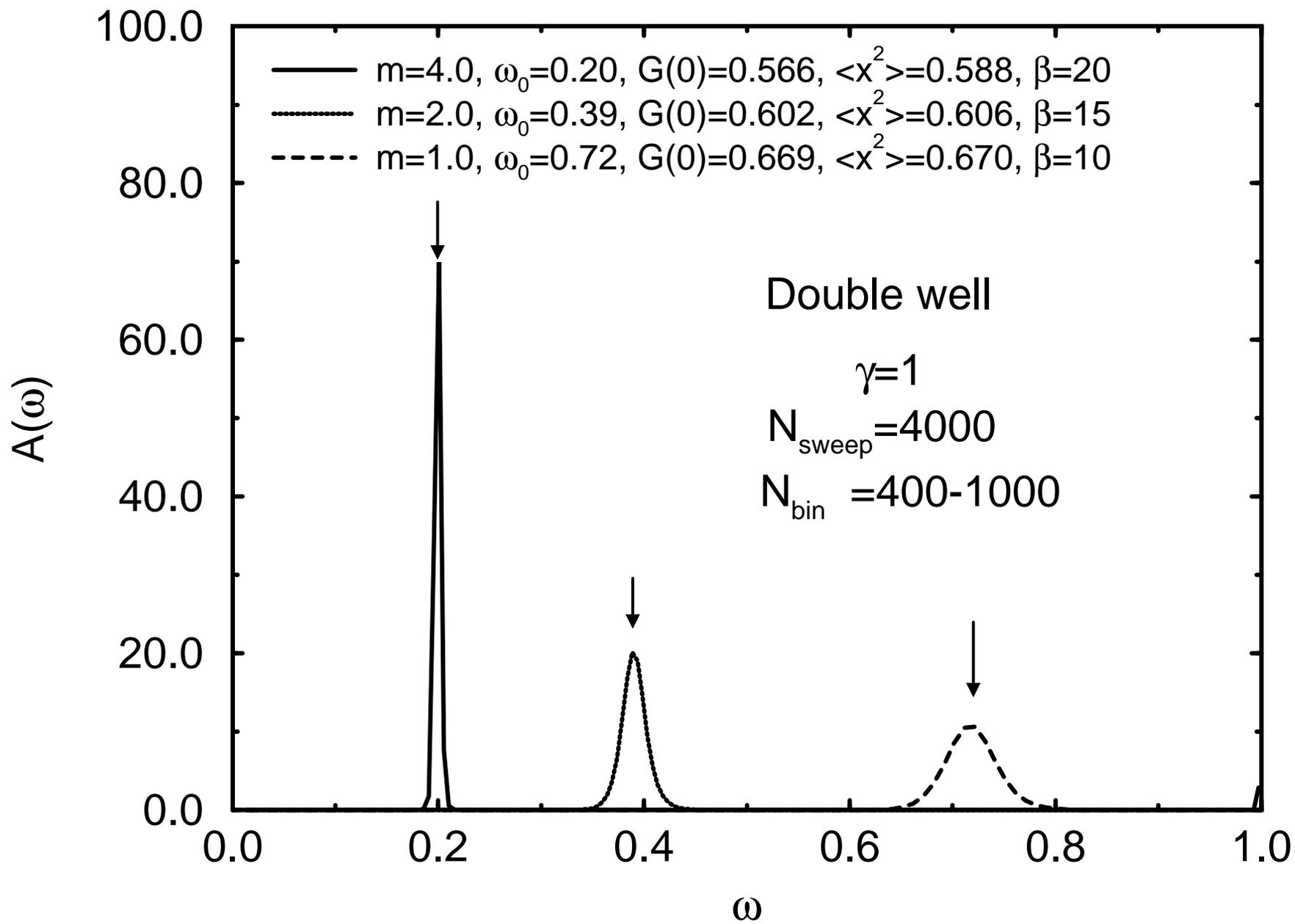
\begin{figure}
\caption[]{Spectral functions for a single double-well potential at
different values of $\gamma$ and different temperatures $1/\beta$.
Arrows above spectral functions denote exact positions $\omega_0$ of
peaks as obtained by numerical solution of the double-well potential.
We also compare sum-rules, calculated by integrating the spectral
functions $A(\omega)$ over $\omega$, with $G(\tau=0)=<x^2>$, obtained
directly from the QMC calculation. $\nu$ is the ratio $1/\eta t_0$
where $\eta$ was obtained by fitting $A(\omega)$ to the analytical
form $G(\omega)$, see Eq.~(23).}
\label{fig8}
\end{figure}

\begin{figure}
\caption[]{Spectral functions $A(\omega)$ for a chain of 20 sites at
two different values of $\epsilon$: (a) $\epsilon = 0.5$ and (b)
$\epsilon = 0.75$. We present the spectral functions for different
wavevectors $k$ in units of $2\pi/20$.}
\label{fig9}
\end{figure}

%
%


\begin{references}
\bibitem[*]{byline} Present Address:  Jo\v zef  Stefan Institute,
University of Ljubljana, 61111 Ljubljana, Slovenia.

\bibitem{berne86} A brief review is given by B.J.\ Berne and D.\
Thirumalai, Ann.\ Rev.\ Chem.\ Phys.\ {\bf 47}, 401 (1986).  More
recent works include: J.D.\ Doll, R.D.\ Coalson, and D.L.\ Freeman,
J.\ Chem.\ Phys.\ {\bf 87}, 1641 (1987): J.D.\ Doll, D.L.\ Freeman,
and T.L.\ Beck, Adv.\ Chem.\ Phys.\ {\bf 78}, 61 (1990); C.H. Mak and
R. Egger, Phys. Rev. A, to appear; and references therein.

\bibitem{gubernatis91} J.E.\ Gubernatis, M.\ Jarrell, R.N.\ Silver, and
D.S.\ Sivia, Phys.\ Rev.\ B {\bf 44}, 6011 (1991); M. Jarrell and J.E.
Gubernatis, Phys. Reprt., to appear.

\bibitem{wang94} Xidi Wang, D.K. Campbell, and J.E. Gubernatis, Phys.
Rev. B {\bf 49}, 15485 (1994).

\bibitem{mahan81} G.D.\ Mahan, {\it Many-Particle Physics\/} (Plenum,
New York, 1981).

\bibitem{creutz81} See for example, M. Creutz and B. Freedman, Ann.
Phys. (N.Y.) {\bf 132}, 427 (1981).

\bibitem{berne??} B.J.\ Berne and D.\ Thirumalai, Ann.\ Rev.\ Phys.\
Chem.\ {\bf 37}, 401 (1986).

\bibitem{duane87} S. Duane, A.D. Kennedy, B.J. Pendleton, and D.
Roweth, Phys. Lett. B {\bf 195}, 216 (1987).

\bibitem{neal95} R.N. Neal, unpublished.

\bibitem{fahy92} S. Fahy and D.R. Hamann, Phys. Rev. Lett. {\bf 69},
761 (1992).

\bibitem{skilling89} J.\ Skilling, in {\it Maximum Entropy and Bayesian
Methods} edited by J.\ Skilling (Kluwer Academic, Dordrecht, 1989), p.\ 45.

\bibitem{gull89} S.F.\ Gull, in {\it Maximum Entropy and Bayesian
Methods} edited by J.\ Skilling (Kluwer, Dordrecht, 1989), p.\ 53.

\bibitem{bryan90} R.K.\ Bryan, Eur.\ Biophys.\ J.\ {\bf 18}, 165
(1990).

\bibitem{frantz92} D.D. Frantz, D.L. Freeman, and J.D. Doll, J. Chem.
Phys. {\bf 97}, 5713 (1992).

\bibitem{cao} Jianshu Cao and B.J.\ Berne, J.\ Chem.\ Phys.\ {\bf 92},
1980 (1990).

\end{references}
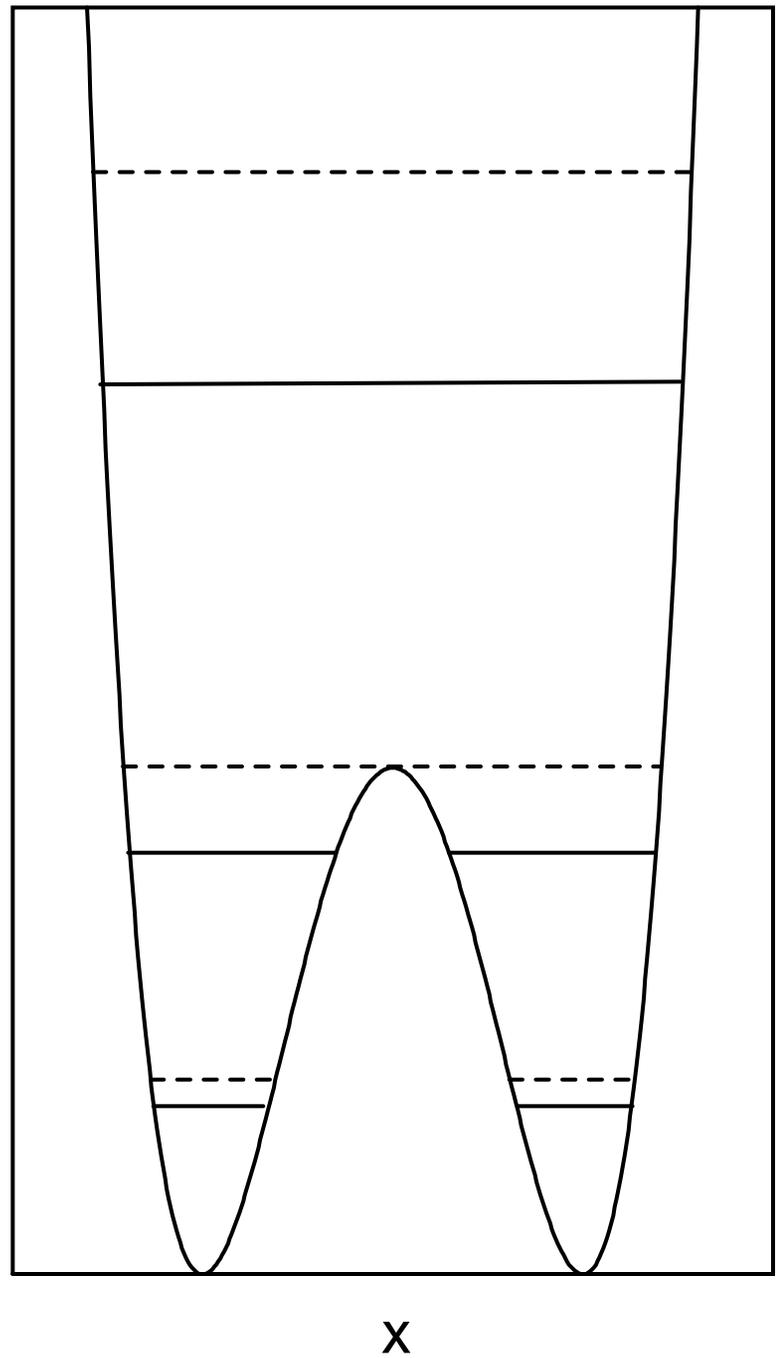
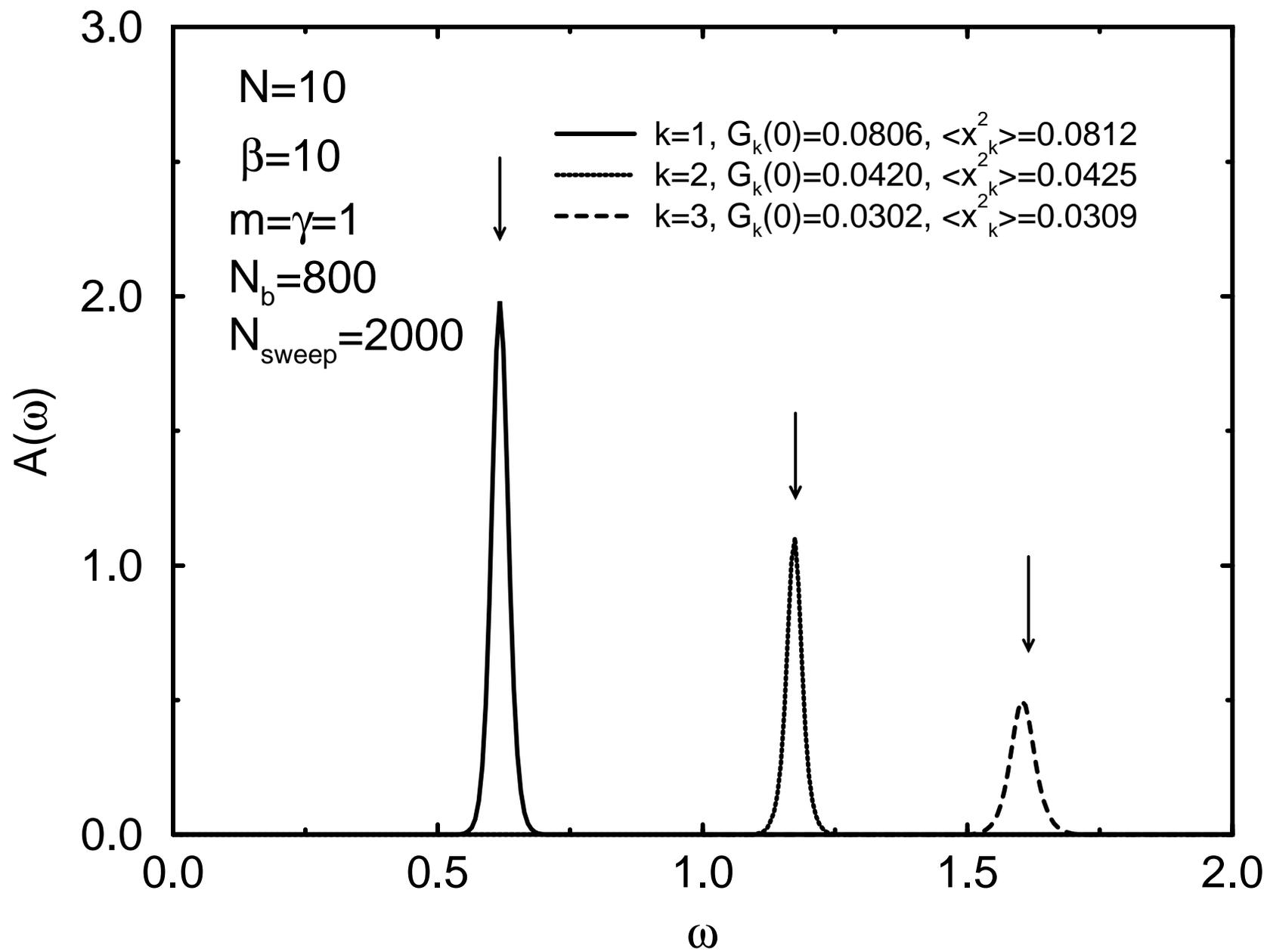
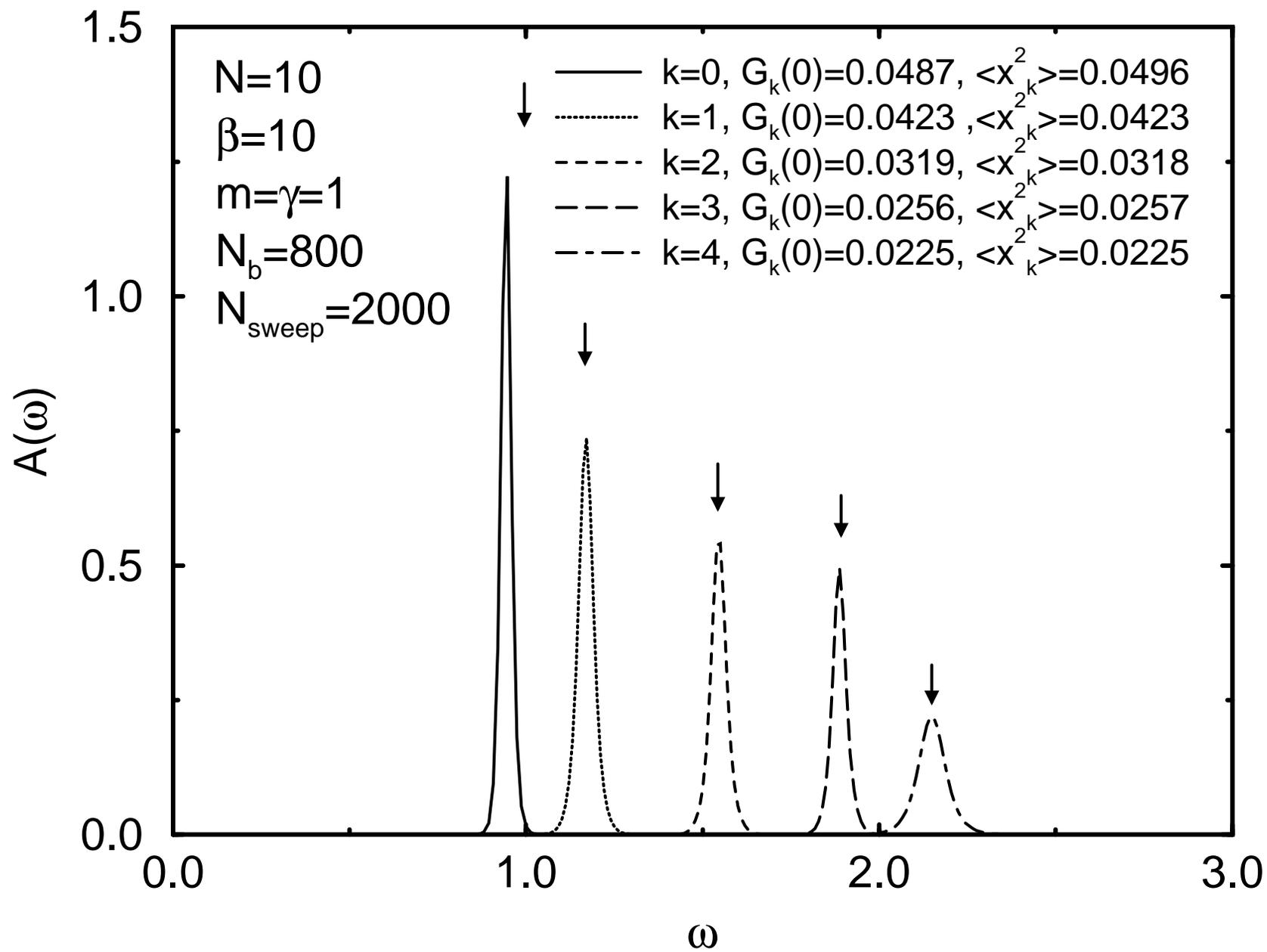
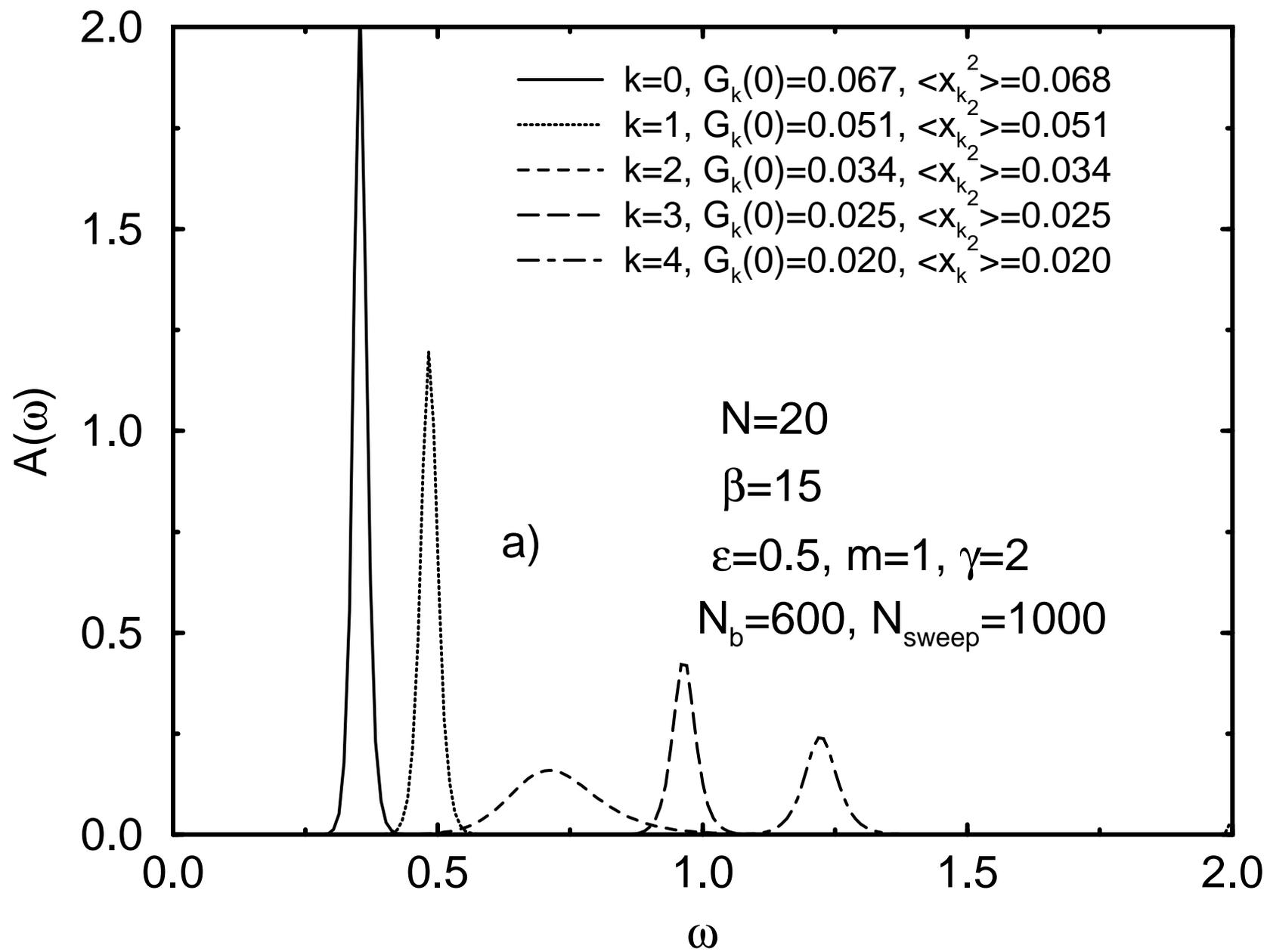
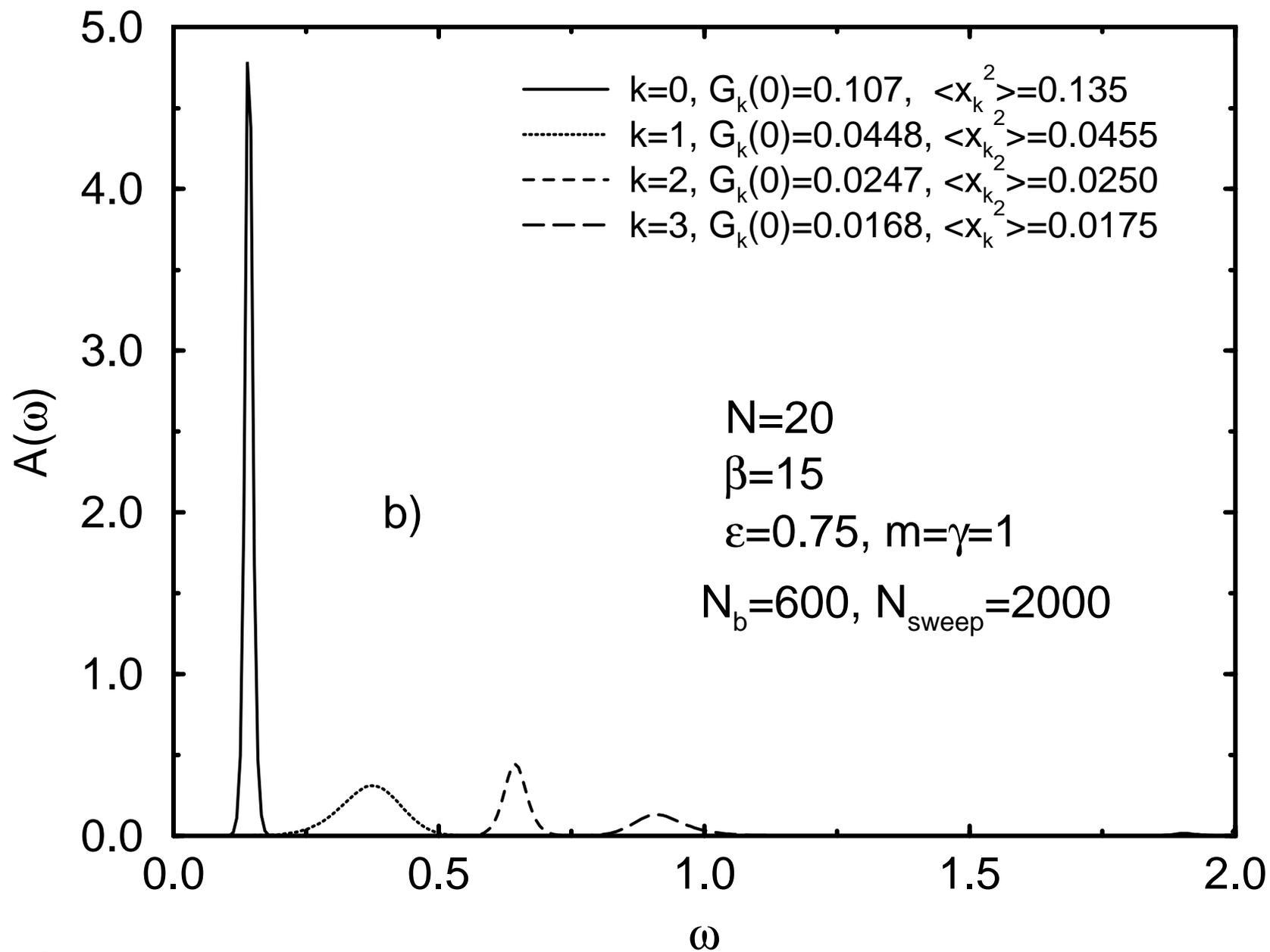
\end{document}